\documentclass[letterpaper, twocolumn, pra]{revtex4}
\usepackage{graphicx, amsmath, amssymb, bm, titlesec,mathtools}

\begin{document}

\title{Generation of a Gaussian Schell-model field as a mixture of its coherent modes}

\author{Abhinandan Bhattacharjee, Rishabh Sahu and Anand K. Jha}

\email{akjha9@gmail.com}

\affiliation{Department of Physics, Indian Institute of
Technology, Kanpur 208016, India}

\begin{abstract}

Gaussian Schell-model fields are examples of spatially partially coherent fields which in recent years have found several unique applications. The existing techniques for generating Gaussian Schell-model (GSM) fields are based on introducing randomness in a spatially completely coherent field and are limited in terms of control and precision with which these fields can be generated. In contrast, we demonstrate an experimental technique that is based on the coherent mode representation of GSM fields. By generating individual coherent eigenmodes using a spatial light modulator and incoherently mixing them in a proportion fixed by their normalized eigenspectrum, we experimentally produce several different GSM fields. Since our technique involves only the incoherent mixing of coherent eigenmodes and does not involve introducing any additional randomness, it provides better control and precision with which GSM fields with a given set of parameters can be generated.

\end{abstract}

\maketitle

\section{Introduction}
Spatially partially coherent fields have been extensively studied in the past few decades, and among such fields, the Gaussian Schell-model (GSM) field has been the most important \cite{mandel1995optical,cai2014josaa,gbur2010els,cai2017els, jha2010pra, collett1978optlet}. This is because GSM fields are widely used in theoretical models due to their simple functional form and have found several unique applications in areas including free space optical communication \cite{ricklin2002josaa, korotkova2004optexp}, ghost imaging \cite{cai2004optlet,cai2005pre}, propagation through random media and atmospheric conditions \cite{gbur2002josaa, wang2016optexp, lee2016ao}, particle trapping \cite{dong2012pra}, and optical scattering \cite{van2010prl}.  A GSM field is characterized by a Gaussian transverse intensity profile and a Gaussian degree of coherence function. The widths of these Gaussian functions are the parameters that characterize a GSM field. There are several experimental techniques for generating GSM fields \cite{de1979optcom,he1988optcom,wang2006optexp,tervonen1992josaa,hyde2015jap}. In all these techniques, partial spatial coherence is generated by introducing randomness in a spatially completely coherent Gaussian beam and then by ensuring that the transverse intensity profile of the randomized field stays Gaussian. The most common way of introducing randomness is by using a rotating ground glass plate (RGGP), which causes the perfect spatial correlation of the incoming field to reduce to a Gaussian correlation. The other ways of introducing randomness include using either an acousto-optic modulator or a spatial light modulator (SLM) \cite{savage2009natphot,gruneisen2008ao,johnson1993ieee}.

Although the above mentioned experimental techniques ensure that the transverse intensity, as well as the degree of coherence of the generated field, are gaussian functions, these techniques are limited in terms of precision and control. Therefore, an efficient experimental technique with precise control for generating a GSM field is still required. In this article, we demonstrate just such a technique, which, in contrast to the techniques mentioned above, does not explicitly involve introducing additional randomness. Our technique is based on the coherent mode representation of GSM fields \cite{starikov1982josaa,gori1980optcom,gori1983optcom}. The coherent mode representation is the way of describing a spatially partially coherent field as a mixture of several spatially completely coherent modes. Therefore, a GSM field with any given set of parameters can be generated with precision by first producing different coherent modes and then incoherently mixing them in a proportion fixed by the coherent mode representation of the field.

\section{Theory}
\subsection{Gaussian Schell-model field as a mixture of its
constituent coherent modes}

The coherent mode representaiton of a one-dimensional GSM field was worked out in Ref.~\cite{starikov1982josaa,gori1980optcom,gori1983optcom}. Here, we present it for the two-dimensional case. The cross spectral density of a GSM field is given by
\begin{equation}
W(\bm{\rho_1},\bm{\rho_2})=\sqrt{I(\bm{\rho_1})I(\bm{\rho_2})}\mu(\bm{\rho_1}-\bm{\rho_2}) \label{CSD_GSM}
\end{equation}
with $I(\bm{\rho_1})=A^2\exp\left[-\rho_1^2/(2\sigma_s^2)\right]$, $ I(\bm{\rho_2})=A^2\exp\left[-\rho_2^2/(2\sigma_s^2)\right]$ and $\mu(\bm{\rho_1}-\bm{\rho_2}) = \exp\left[-(\Delta\rho)^2/(2\sigma_g^2)\right]$. Here, $\bm{\rho_1}\equiv (x_1, y_1)$, $\bm{\rho_2}\equiv (x_2, y_2)$, $\rho_1=|\bm{\rho_1}|$, $\rho_2=|\bm{\rho_2}|$, $\Delta\rho=|\bm{\rho_1}-\bm{\rho_2}|$, and $A$ is a constant. $I(\bm{\rho_1})$ is the intensity at point $\bm{\rho_1}$ with $\sigma_s$ being the r.m.s width of the beam, and $\mu(\bm{\rho_1}-\bm{\rho_2})$ is the degree of spatial coherence between points $\bm{\rho_1}$ and $\bm{\rho_2}$, with $\sigma_g$ being the r.m.s spatial coherence width of the beam. $W(\bm{\rho_1}, \bm{\rho_2})$ can be written in terms of its coherent mode representation as \cite{starikov1982josaa}: $
W(\bm{\rho_1}, \bm{\rho_2})=\sum_m \sum_n \lambda_{mn} W_{mn}(\bm{\rho_1}, \bm{\rho_2})$,
where $\lambda_{mn}$ are the eigenvalues and $W_{mn}(\bm{\rho_1}, \bm{\rho_2})=\phi^*_{mn}(\bm{\rho_1})\phi_{mn}(\bm{\rho_2})$ are the cross-spectral density functions corresponding to the coherent eigenmodes $\phi_{mn}(\bm{\rho})$. 
\begin{figure*}[t!]
\begin{center}
\includegraphics{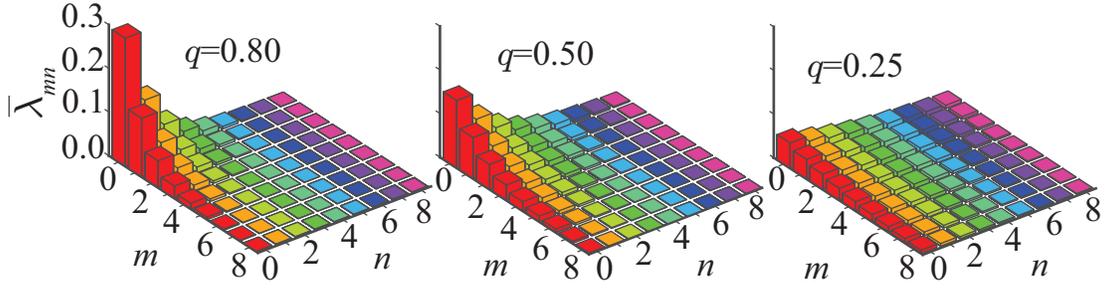}
\caption{Theoretical plots of the normalized eigenvalues $\bar\lambda_{mn}$ for three different values of the degree of global coherence, namely, for $q=0.80$, $q=0.50$, and $q=0.25$. }\label{fig1}
\end{center}
\end{figure*}
Following Ref.~\cite{starikov1982josaa}, we write Eq.~(\ref{CSD_GSM}) as 
\begin{equation}
W(\bm{\rho_1},\bm{\rho_2})=\sum_m \sum_n \lambda_{mn} \phi^*_{mn}(x_1, y_1)\phi_{mn}(x_2, y_2) \label{CMR_GSM2}
\end{equation}
where
\begin{eqnarray}
&\lambda_{mn}=A^2 \left(\frac{\pi}{a+b+c}\right)\left(\frac{b}{a+b+c}\right)^{(m+n)} \nonumber \\
& \phi_{mn}(x,y)=\left( \frac{2c}{\pi} \right)^\frac{1}{2} \frac{1}{\sqrt{2^{m+n} m!n!}} H_m(x\sqrt{2c}) \notag \\ & H_n(y\sqrt{2c})e^{-c(x^2+y^2)} \nonumber
\end{eqnarray}
We have $a=1/(4\sigma_s^2),  b=1/(2\sigma_g^2),  c=(a^2+2ab)^{1/2}$, and $H_m(x)$ are the Hermite polynomials. Substituting $\sigma_g/ \sigma_s=q$, we write $\lambda_{mn}$ as
\begin{equation}
\lambda_{mn}=\lambda_{00}\left[ \frac{1}{(q^2/2)+1 +q[(q/2)^2+1]^{1/2}}\right]^{(m+n)}\label{eigenvalue}
\end{equation}

%
The quantity $q$ is a measure of the ``global degree of coherence" of the field. For fixed $\sigma_s$, higher values of $q$ imply higher values for the degree of spatial coherence. In what follows it will be very convenient to work with the normalized eigenvalues $\bar\lambda_{mn}$. So, for that purpose, we first take $\lambda_{00}=1$ and then define $\bar\lambda_{mn}$ as :$
\bar\lambda_{mn}=\lambda_{mn}/\left(\sum_{mn}\lambda_{mn}\right)$
such that $\sum_{mn}\bar\lambda_{mn}=1$. 

The coherent mode representation describes a partially coherent field as an incoherent mixture of completely uncorrelated coherent modes. We note that Eqs.~(\ref{CSD_GSM}) and (\ref{CMR_GSM2}) are the two equivalent descriptions of the same cross-spectral density function $W(\bm{\rho_1}, \bm{\rho_2})$ for the GSM field. While Eq.~(\ref{CSD_GSM}) describes $W(\bm{\rho_1}, \bm{\rho_2})$ as a single partially coherent field, Eq.~(\ref{CMR_GSM2}) describes it as an incoherent mixture of spatially completely coherent eigenmodes  $\phi_{mn}(x, y)$ with their proportions given by the normalized eigenvalues $\bar\lambda_{mn}$. The intrinsic randomness of the GSM field, which is described by Eq.~(\ref{CSD_GSM}) as being across the transverse plane of the field, gets described by Eq.~(\ref{CMR_GSM2}) as complete randomness between different coherent eigenmodes. Equation~(\ref{CMR_GSM2}) shows that to generate a GSM field, one needs to generate the spatially completely coherent eigenmodes $\phi_{mn}(x, y)$ and then mix them incoherently in $\bar\lambda_{mn}$ proportion. We also find that for a normalized eigenspectrum, the coherent mode representation of Eq.~(\ref{CMR_GSM2}) has only $q$ and $c$ as free parameters.  The parameter $q$ decides the exact proportion $\bar\lambda_{mn}$  of the eigenmodes $\phi_{mn}(x, y)$ and the parameter $c$ decides the overall transverse extent of the field. Thus by controlling  $q$ and $c$ one can generate any desired GSM field.

The coherent mode representation of a GSM field having only one term represents a completely coherent Gaussian field with $\sigma_g=\infty$. On the other hand, a completely incoherent GSM field implies $\sigma_g\to 0$, and in this limit, the coherent mode representation contains an infinite number of terms. When $\sigma_g$ is finite, the field is partially coherent, and the coherent mode representation contains only a finite number of terms with significant eigenvalues. Figure \ref{fig1} shows the theoretical plots of normalized eigenvalues $\bar\lambda_{mn}$  for three different values of $q$, namely, $q=0.8$, $q=0.5$, and $q=0.25$.  The value of $c$ for all the fields is 1.34 mm$^{-2}$.  We find that to generate GSM fields with the smaller global degree of coherence $q$ one requires to mix a larger number of eigenmodes.

\subsection{Measuring Gaussian Schell-model fields}
For measuring the cross-spectral density of a Gaussian Schell-model field, we use the measurement technique of Ref.~\cite{bhattacharjee2018apl}. This is a two-shot technique involving wavefront-inversion inside an interferomter and is the spatial analog of the technique \cite{kulkarni2017natcomm} recently demonstrated for measuring the angular coherence function \cite{jha2011pra}. Figure \ref{fig2}(b) shows the schematic diagram of the measurement technique. For a GSM field input, the intensity $I_{\rm out}(\bm\rho)$ at the output port of the interferometer is given by $I_{\rm out}(\bm\rho)=k_1I(\bm\rho) + k_2 I(-\bm\rho)+ 2\sqrt{k_1k_2}W(\bm\rho, -\bm\rho)\cos\delta$  \cite{bhattacharjee2018apl}.
Here $k_1$ and $k_2$ are the scaling constants in the two arms and  $\delta$ is the overall phase difference between the two interferometric arms; $I(\bm\rho)$ is the intensity of the GSM field at point $\bm\rho$ and $W(\bm\rho, -\bm\rho)$ is the cross spectral density of the GSM field for the pair of points $\bm\rho$ and $-\bm\rho$. Now, suppose there are two output interferograms with intensities $\bar{I}^{\delta_c}_{\rm out}(\bm\rho)$ and 
$\bar{I}^{\delta_d}_{\rm out}(\bm\rho)$ measured at $\delta=\delta_c$ and $\delta=\delta_d$, respectively. As worked out in Ref.~\cite{bhattacharjee2018apl}, if the shot-to-shot variation in the background intensity is negligible, the difference $\Delta\bar{I}_{\rm out}(\bm\rho)=\bar{I}^{\delta_c}_{\rm out}(\bm\rho)-\bar{I}^{\delta_d}_{\rm out}(\bm\rho)$ in the intensities of the two interferograms is given by $
\Delta\bar{I}_{\rm out}(\bm\rho)=2\sqrt{k_1k_2} (\cos\delta_c-\cos\delta_d)W(\bm\rho, -\bm\rho) $. We find that the difference intensity is proportional to the cross-spectral density function $W(\bm\rho, -\bm\rho)$. Using Eq.~(\ref{CSD_GSM}), we write $
W(\bm\rho, -\bm\rho) \rightarrow W(2\bm\rho)=W(2x, 2y)
=A^2\exp\left[-\frac{(2\rho)^2}{8\sigma_s^2}\right] \exp\left[-\frac{(2\rho)^2}{2\sigma_g^2}\right]$, and therefore we get
%

\begin{equation}
\Delta\bar{I}_{\rm out}(\bm\rho)\propto W(2\bm\rho) \label{diff_Iout}
\end{equation}
that is, the difference intensity $\Delta\bar{I}_{\rm out}(\bm\rho)$ is proportional to $W(2\bm\rho)$. Thus by measuring the difference intensity the cross-spectral density function $W(2\bm\rho)$ can be directly measured without having to know $k_1,k_2$ and $\delta$ precisely. Using $W(2\rho)$ and $I(\rho)$ , the degree of coherence $\mu(2\rho)$ of the field can be written as

\begin{equation}
\mu(2\bm{\rho})=\frac{W(2\bm{\rho})}{\sqrt{I(\bm{\rho})I(-\bm{\rho})}}=\frac{W(2\bm{\rho})}{I(\bm{\rho})}\label{coh_mu} 
\end{equation}

where $I(\bm{\rho})=I(-\bm{\rho})$ because the transverse intensity profile of a GSM field is symmetric about inversion. 

\section{EXPERIMENTAL RESULTS}
\subsection{Generation}
\begin{figure}[t!]
\begin{center}
\includegraphics{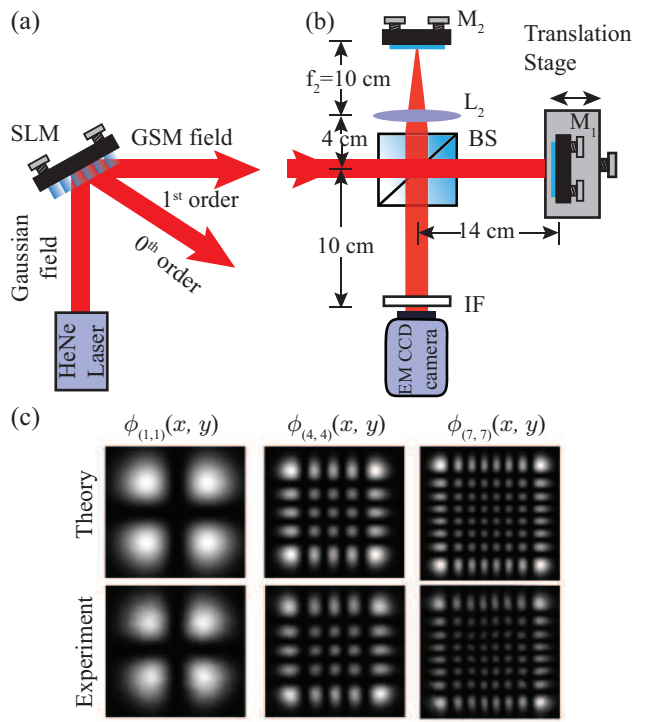}
\caption{(a) Schematic setup for generating GSM fields. (b) Schematic setup for measuring the cross-spectral density function using the technique of Ref.~\cite{bhattacharjee2018apl}. Here, we have SLM: Spatial Light Modulator; BS: beam splitter; M: mirror; and L: converging lens. (c) The theoretically expected and experimentally generated intensity corresponding to the eigenmodes $\phi_{11}(x, y)$, $\phi_{44}(x, y)$, and $\phi_{77}(x, y)$. }\label{fig2}
\end{center}
\end{figure}

\begin{figure*}[t!]
\begin{center}
\includegraphics{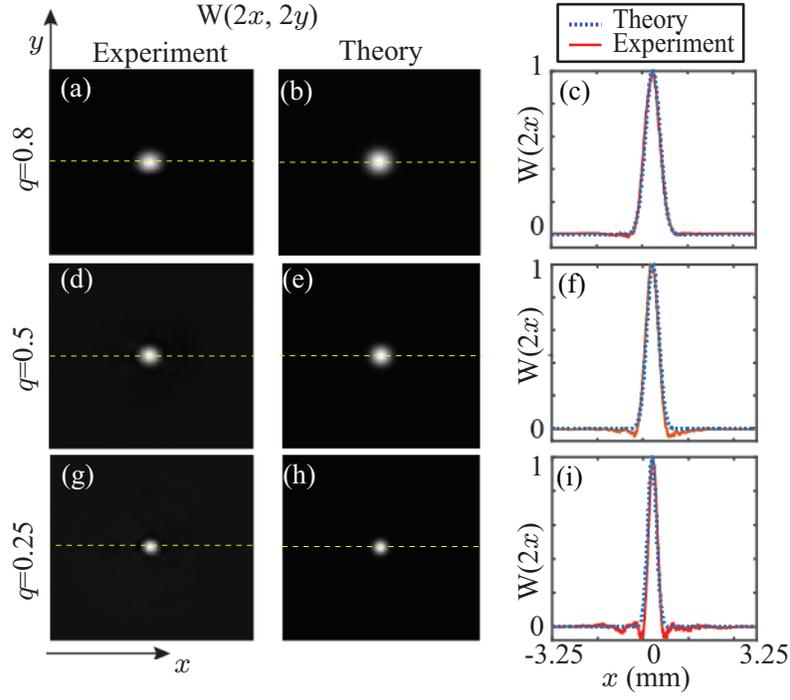}
\caption{
Plots of the the cross-spectral density function of GSM fields with $q=0.8$, $q=0.5$,and $q=0.25$.
For the three values of $q$,
(a),(d) and (g) are the experimentally measured cross-spectral density functions $W(2x, 2y)$ while (b),(e) and (h) are the corresponding theoretical plots. (c),(f) and (i) are the plots of the one-dimensional cuts along the $x$-direction of the theoretical and experimental cross-spectral density functions.
 }\label{fig3}
\end{center} 
\end{figure*}
Figures \ref{fig2}(a) and \ref{fig2}(b) show the experimental setup for generating the GSM field and measuring its cross-spectral density function in a two-shot manner \cite{bhattacharjee2018apl}, respectively. The Gaussian field from a 5-mW He-Ne laser incident on a Holoeye Pluto spatial light modulator (SLM) and an appropriate phase pattern is displayed on the SLM to generate a given eigenmode at the detection plane of the EMCCD camera. In particular, the SLM is programmed to generate different eigenmodes $\phi_{mn}(x, y)$ using the Arrizon method \cite{arrizon2007josaa}. Figure \ref{fig2} (c) shows the experimentally measured and theoretically expected intensity profiles of eigenmodes: $\phi_{11}(x, y)$, $\phi_{44}(x, y)$, and $\phi_{77}(x, y)$. We find a good match between the theory and experiment. Now, to produce a GSM field with a given $q$, that is, a given eigenspectrum $\bar\lambda_{mn}$, we need to produce the incoherent mixture of different eigenmodes with proportion given by $\bar\lambda_{mn}$. It has done in the following manner. First, the phase patterns corresponding to different eigenmodes are displayed on the SLM sequentially. The weights $\bar\lambda_{mn}$ are fixed by making the display-time of the phase pattern corresponding to an eigenmode $\phi_{mn}(x, y)$ proportional to the corresponding eigenvalue $\bar\lambda_{mn}$. In our experiment, the SLM works at 60 Hz. The display-time of a given phase pattern on the SLM is of the order of tens of milliseconds while the coherence time of our He-Ne laser is in tens of picoseconds. Although the SLM introduces a deterministic phase modulation along the beam cross-section for a given eigenmode, the phase modulation for a given eigenmode is completely uncorrelated with that for any other eigenmode. In this way, the SLM produces an incoherent mixture of coherent modes, as long as the time of observation is kept long enough for all the modes to get detected. Therefore, the exposure time of the EMCCD camera is made equal to the total display-time of all the phase patterns such that the camera collects all the generated eigenmodes. 

Using the procedure described above, we generate GSM fields for three different values of $q$, namely, $q=0.8$, $q=0.5$, and $q=0.25$. Although in principle for any given $q$ we need an infinite number of modes to produce the corresponding GSM field precisely. However, the plots in Fig.~\ref{fig1} show that for any finite $q$ the number of eigenvalues $\bar\lambda_{mn}$ with significant contributions are only finite and that the number of significant eigenvalues increases with decreasing $q$. In our experiment, we keep $0.07 \times\bar\lambda_{00}$ as the cutoff for deciding the eigenmodes with a significant contribution. This means that for a given $q$ we generate only those eigenmodes for which $\bar\lambda_{mn} \geq 0.07 \times\bar\lambda_{00}$. With this cutoff, we generate 10, 21, and 66 eigenmodes, respectively, for the three values of $q$. The sum of these eigenvalues $\sum_{mn}\bar\lambda_{mn}$ turns out to be about 0.87, 0.84, and 0.82, respectively, for the three $q$ values, which are quite close to one. 

\begin{figure*}[t!]
\begin{center}
\includegraphics{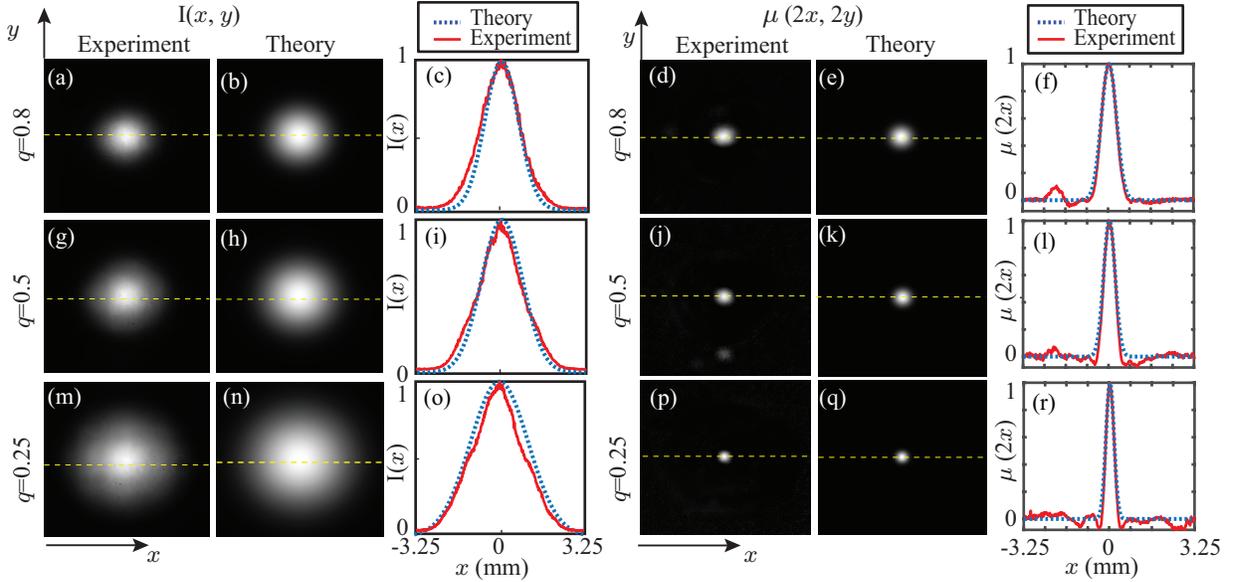}
\caption{Plots of the intensity and the degree of coherence of GSM fields with $q=0.8$, $q=0.5$,and $q=0.25$.  For the three values of $q$, (a),(g) and (m) show the experimentally measured intensity profiles $I(x, y)$ while (b),(h) and (n) are the corresponding theoretical plots. (c),(i) and (o) are the plots of the one-dimensional cuts along the $x$-direction of the theoretical and experimental intensity profiles. For the three values of $q$,
(d),(j) and (p) are the experimental degree of coherence  $\mu(2x, 2y)$ while (e),(k) and (q) are the corresponding theoretical plots. (f),(l) and (r) are the plots of the one-dimensional cuts along the $x$-direction of the theoretical and experimental degree of coherence functions.
 }\label{fig4}
\end{center} 
\end{figure*}

\subsection{Measurement} 
Each of the generated GSM fields is made incident on the interferometer in Figure ~\ref{fig2}(b). In our experiment, the SLM works at 60 Hz, and the EMCCD camera was kept opened for 1.40, 3.00, and 5.76 seconds, respectively, for the three $q$ values. This was for ensuring that the  camera collects all the generated eigenmodes. The value of $c$ in each case was 1.34 mm$^{-2}$. Now, we measure the cross-spectral density function. For each value of $q$, we collect two interferograms, one with $\delta=\delta_c\approx 0$ and the other one with $\delta=\delta_d\approx \pi$. These two interferograms are then subtracted to generate the difference intensity $\Delta\bar{I}_{\rm out}(\bm\rho)$, which is then scaled such that the value of the most intense pixel is equal to one. From equation ~(\ref{diff_Iout}), we have that the scaled difference intensity $\Delta\bar{I}_{\rm out}(\bm\rho)$ is nothing but the scaled cross-spectral density function $W(2\bm\rho)=W(2x, 2y)$. To ensure that the interferograms are not drifting and that the shot-to-shot background intensity variation is negligible, we cover the entire interferometer with a box. Figures \ref{fig3}(a), \ref{fig3}(d), and \ref{fig3}(g) show the experimentally measured cross-spectral density functions for the three values of $q$ while  Figs.~\ref{fig3}(b), \ref{fig3}(e), and \ref{fig3}(h) show the corresponding theoretical cross-spectral density functions
plotted using equation ~(\ref{CSD_GSM}). In order to compare our experimental results with the theory, we take the one-dimensional cuts of the theoretical and experimental cross-spectral density functions and plot them together in Figs.~\ref{fig3}(c), \ref{fig3}(f), and \ref{fig3}(i), for the three values of $q$. 


Next, we measure the transverse intensity profile of the GSM field for different values of q. For measuring the intensity profile, we block the interferometric arm containing the lens and record the intensity at the EMCCD camera plane. Figures \ref{fig4}(a), \ref{fig4}(g) and \ref{fig4}(m) show the measured intensity profiles $I(\bm\rho)=I(x, y)$ for the three values of $q$. The corresponding theoretical intensities as given by equation ~(\ref{CSD_GSM}) are plotted in Figures ~\ref{fig4}(b), \ref{fig4}(h) and \ref{fig4}(n), respectively. The experimental and theoretical plots are both scaled such that the value of the most intense pixel is equal to one. Again, for comparing our experimental results with theory, we plot in Figures ~\ref{fig4}(c), \ref{fig4}(i) and \ref{fig4}(o) the one-dimensional cuts along the $x$-direction of the theoretical and experimental intensity profiles. 

Finally, using the intensity and cross-spectral density measurement above, we find the degree of coherence
$\mu(2x,2y)$. Figures ~\ref{fig4}(d), ~\ref{fig4}(j) and ~\ref{fig4}(p) show the experimental degree of coherence for the three $q$ values  while  Figs.~\ref{fig4}(e), \ref{fig4}(k), and \ref{fig4}(q) show the corresponding theoretical plots. We scale the both experimental and theoretical plot such that the value of the most intense pixel is equal to one. To further compare our experimental results with the theory, we take the one-dimensional cuts along the $x$-direction of the theoretical and experimental degree of coherence and plot them together in Figs.~\ref{fig4}(f), \ref{fig4}(l) and \ref{fig4}(r) for three values of q. The results show that with decreasing $q$ the width of the degree of coherence decreases while the width of the transverse intensity profile increases. This is due to the fact that for generating field with large $q$ values, one requires to mix
together a lower number of eigenmodes, whereas for generating fields with smaller q values, one is required to mix together a larger number of eigenmodes, as illustrated in figure ~\ref{fig1}. We note that although the individual eigenmodes are perfectly spatially coherent,  the increase in the number of eigenmodes in the incoherent mixture increases the randomness and thereby decreases the width of the degree of coherence function, while increasing the transverse width of the beam. We find a good match between the theory and experiment for each value of $q$.

From the above result, it is evident that the GSM field with $q=0.80$  has a better match with the theory than the field with $q=0.25$. The reasons for this are as follows. First of all, as
illustrated in figure ~\ref{fig1}, the eigenvalue distribution for $q=0.25$ is much broader than that for $q=0.80$. 
\begin{figure*}[t!]
\begin{center}
\includegraphics{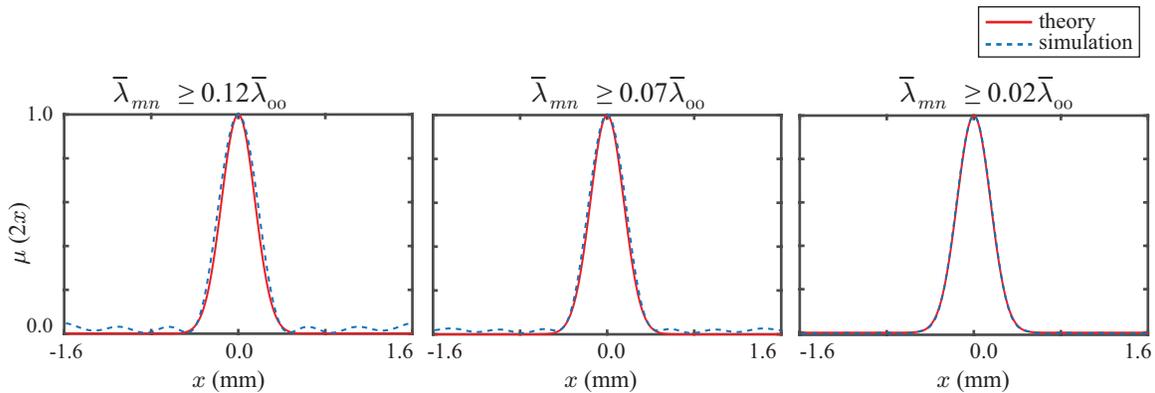}
\caption{Plots of the numerically simulated degree of coherence function for $q = 0.25$ have been shown for the three values of the cutoff on
$\bar\lambda_{mn}$. For each plot, the solid line represents the theoretical degree of coherence generated using equation~\ref{CSD_GSM}.
 }\label{fig5}
\end{center} 
\end{figure*}
As a result, for producing the GSM field with $q = 0.25$, we need to generate a larger number of modes with corresponding eigenvalues $\bar\lambda_{mn}$.
As mentioned earlier, the eigenvalue$\bar\lambda_{mn}$ is assigned by the display time of the corresponding eigenmode $\phi_{m,n}(x,y)$ on the SLM. Now, since the refresh rate of the SLM is 60 Hz, and the collection time of the detection camera is in seconds, we have only a few hundred discrete time-bins for assigning the eigenvalues $\bar\lambda_{mn}$. This puts a limit on the precision with which a given number of modes with
eigenvalues $\bar\lambda_{mn}$ could be generated and therefore results in a
better match for GSM field with $q = 0.80$ since that requires a
lower number of modes to be produced. Nevertheless, this
limitation can be overcome by using a faster SLM, which can
provide a greater number of time-bins for a given collection
time and thereby can improve the precision with which $\bar\lambda_{mn}$
could be generated. The other reason for a better match at
$q = 0.80$ is the cutoff on $\bar\lambda_{mn}$, which restricts the number of eigenmodes in the incoherent mixture. For the given cutoff of
$0.07 \times\bar\lambda_{00}$ on $\bar\lambda_{mn}$, the sum $\sum_{mn}\bar\lambda_{mn}$ becomes 0.87 and 0.82 for
$q = 0.80$ and $0.25$, respectively, resulting in a better match for
$q = 0.80$. Figure.~\ref{fig5} shows the effect of the cutoff on the precision with which GSM field could be generated. In the figure,
we have plotted the numerically simulated degree of coherence
for $q = 0.25$ for three different values of the cutoffs. We see that
as the cutoff is lowered, a GSM field with a better match with
the theory can be produced. Thus, in our technique, one can
decide the cutoff for the eigenvalues depending on the precision
requirement for a given application.

\section{Discussion and Conclusion}
In conclusion, we have demonstrated in this article an experimental technique for generating GSM fields that is based on its coherent mode representation. We have reported the generation of GSM fields with a range of values for the global degree of coherence, and to the best of our knowledge, such a
demonstration has not been reported earlier. Compared to the
existing techniques for producing GSM fields \cite{de1979optcom,he1988optcom,wang2006optexp,tervonen1992josaa,hyde2015jap}, the main advantage of our technique is that it does not explicitly involve
introducing any additional randomness. As a result, the errors
involved in our scheme are mostly systematic and are easily
controllable. This fact has also been highlighted in the recently
demonstrated techniques \cite{aarav2017pra,chen2018ol,rodenburg2014josab}  for producing different types
of spatially partially coherent fields without introducing additional randomness. The other advantage of our method is that it
is SLM-based; therefore the phase patterns can be changed
electronically and different GSM fields can be generated by just
controlling c and q without having to remove any physical
element from a given experimental setup. This is as opposed to
when producing a partially coherent field using an RGGP, in
which case one requires separate RGGPs with precisely characterized features for producing different GSM fields. Thus, our
method provides much better control and precision with which
GSM fields can be generated. We therefore expect our method
to have important practical implications for applications that are
based on utilizing spatially partially coherent fields

\section{Acknowledgement}
We thank Shaurya Aarav for useful discussions and acknowledge financial support through the research
grant no. EMR/2015/001931 from the Science
and Engineering Research Board (SERB), Department
of Science and Technology, Government of India.


%

\end{document}